\documentclass[twocolumn,prl,10pt,amsmath,amssymb,nofootinbib,showpacs,superscriptaddress,floatfix]{revtex4-1}

\usepackage{graphicx}
\usepackage{dcolumn}
\usepackage{bm}

\usepackage{color}
\usepackage[usenames,dvipsnames]{xcolor}
\usepackage[colorlinks=true,linkcolor=Red,citecolor=Green,linktoc=page]{hyperref}

\newcommand{\rs}{\rm\scriptscriptstyle}

\begin{document}


\title{The Superconductor-Superinsulator Transition: \\ S-duality and the QCD on the desktop}

\author{M. Cristina Diamantini}
\affiliation{
NiPS Laboratory, INFN and Dipartimento di Fisica, University of Perugia, via A. Pascoli, I-06100 Perugia, Italy
}

\author{Luca Gammaitoni}
\affiliation{
NiPS Laboratory, INFN and Dipartimento di Fisica, University of Perugia, via A. Pascoli, I-06100 Perugia, Italy
}

\author{Carlo A. Trugenberger}
\affiliation{
SwissScientific, rue due Rhone 59, CH-1204 Geneva, Switzerland
}

\author{Valerii M. Vinokur}
\affiliation{Materials Science Division, Argonne National Laboratory, Argonne, Illinois 60439, USA}


\begin{abstract}
We show that the nature of quantum phases around the superconductor-insulator transition (SIT) is controlled by charge-vortex topological interactions, and does not depend on the details of material parameters and disorder. We find three distinct phases, superconductor, superinsulator and bosonic topological insulator. The superinsulator is a state of matter with infinite resistance in a finite temperatures range, which is the S-dual of the superconductor and in which charge transport is prevented by electric strings binding charges of opposite sign. The electric strings ensuring linear confinement of charges are generated by instantons and are dual to superconducting Abrikosov vortices. Material parameters and disorder enter the London penetration depth of the superconductor, the string tension of the superinsulator and the quantum fluctuation parameter driving the transition between them. They are entirely encoded in four phenomenological parameters of a topological gauge theory of the SIT. Finally, we point out that, in the context of strong coupling gauge theories, the many-body localization phenomenon that is often referred to as an underlying mechanism for superinsulation is a mere transcription of the well-known phenomenon of confinement into solid state physics language and is entirely driven by endogenous disorder embodied by instantons with no need of exogenous disorder.

\end{abstract}

\maketitle


The superconductor-insulator transition (SIT)\,\cite{efetov, haviland, paalanen, fisher1990, fisher1992, fazio, goldman} is a paradigmatic quantum phase transition found in Josephson junction arrays (JJA)\,\cite{efetov,fazio} and in 2D disordered superconducting films at low temperatures $T$\,\cite{haviland,paalanen,fisher1990,fisher1992}. The tuning parameter driving the SIT is the ratio of the single junction Coulomb energy  to the Josephson coupling. In 2D films this ratio is effectively controlled by varying the film thickness $d$ which regulates the strength of disorder and hence of Coulomb screening or by applying a magnetic field that suppresses the Josephson coupling. This can cause a dramatic change in the ground state so that superconductivity is lost in favour of insulating behaviour. 

In 1978, G. 't Hooft\,\cite{thooft} appealed to a solid state physics analogy in a Gedankenexperiment to explain quark confinement and demonstrated that this is realized in a phase which is in many respects similar to the superconducting phase, but is in a sense a zero particle mobility phase, the extreme opposite of a superconductor and called hence this phase a "superinsulator." In 1996, two of the present authors (mcd and cat)\,\cite{dst} developed a comprehensive field theory framework for the description of the SIT in JJA. They predicted that, on the insulating side of the SIT, a new ground state forms, corresponding to a novel phase with infinite resistance. This novel phase is dual to the superconductor, characterized by zero resistance, and they thus independently also called this phase a superinsulator. 
Independently, superinsulators where also soon proposed in\,\cite{doniach1998}. Finally, the name and phenomenon of superinsulation was rediscovered and experimentally detected by one of the authors (vmv) and his collaborators in\,\cite{vinokur2008superinsulator, vinokurannals} based on the earlier experimental observations\,\cite{Shahar2005,Baturina2007}. Superinsulators where identified in\,\cite{vinokur2008superinsulator} as a low-temperature charge Berezinskii-Kosterlitz-Thouless (BKT)\cite{ber,KT} phase emerging at the temperature of the charge BKT transition and was derived from the wave function -- phase-amplitude duality of the uncertainty principle. 

Motivated by 't Hooft's beautiful idea\,\cite{thooft} and building 
on the framework proposed in\,\cite{dst}, a comprehensive theory of the SIT and superinsulators was developed in\,\cite{dtv}. It was shown that in duality to the Meissner effect in superconductors, which constricts the magnetic field lines penetrating a type II superconductor into Abrikosov vortices, in superinsulators, electric flux tubes that linearly bind Cooper pairs into neutral ``mesons" form. These electric strings confine fluctuating $\pm$Cooper pair charges completely, thereby impeding electric conductance: Cooper pairs are confined exactly as quarks in hadrons\,\cite{dtv}, the finite-temperature deconfinement transition for linear potentials coinciding with the 2D BKT transition\,\cite{yaffe, vft}. This established superinsulators as a novel, distinct state of matter.  

The interpretation of the infinite-resistance state in terms of the superinsulating state dual to superconductivity was fiercely criticized in\,\cite{critique} by questioning the correctness of the microscopic modeling of the superconducting films as a Josephson junction array (JJA) and attacking the treatment of this array by\,\cite{vinokur2008superinsulator}. Here we demonstrate by a straightforward calculation that the nature of the phases in the critical vicinity of the SIT is determined solely by fundamental topological interactions and by gauge invariance. Disorder plays only the role of the tuning mechanism cranking up and down the effective strength of the Coulomb interaction that drives the system across the SIT. Our finding relates the topological nature of the superinsulator with the earlier finding\,\cite{vinokurannals} that superinsulators, dual to superconductors, emerge as a consequence of quantum conjugation of the number of particles $N$ and phase $\varphi$ in the Cooper pair condensate, $[N,\varphi]=i$, and the related competition of the uncertainties of these variables due to the Heisenberg principle, $\Delta\varphi\Delta N\geqslant 1$. This completely invalidates the critique\,\cite{critique} since microscopic details of the model are irrelevant. 

The $T=0$ partition function of a quantum many body system in $D$ spatial dimension is determined by the Euclidean action of a classical system in $(D+1)$ dimensions. The effective action governing this partition function can then be expanded in a series of derivatives. The universal properties of the system, including the phase structure and the nature of the possible phases are determined by the relevant and marginal terms in this effective action. The more terms one includes, the more microscopic details are modeled and taken into account. 

It has been recognized since the very early days of the SIT studies, that the relevant degrees of freedom for the SIT are charges (Cooper pairs) and vortices and that the possible phases are determined by the competition between these two types of degrees of freedom \cite{fisher1990,fisher1992,fazio}. 
Charges and vortices are subject to topological interactions, embodied by the Aharonov-Bohm/Aharonov-Casher (ABC) phases they acquire when  encircling one another. A local formulation of such topological interactions requires the introduction of two emergent gauge fields $a_{\mu }$ and $b_{\mu}$ coupled to the conserved charge and vortex currents, respectively. The Euclidean topological action for these gauge fields is the mixed Chern-Simons action \cite{jackiw} 
\begin{equation} 
S =  \int d^3x \ i  {\kappa \over 2\pi} a_{\mu} \epsilon_{\mu \alpha \nu} \partial_{\alpha} b_{\nu} + i\sqrt{\kappa} a_{\mu} Q_{\mu} +i\sqrt{\kappa} b_{\mu} M_{\mu} \  ,
\label{CS}
\end{equation}
where $\kappa$ is the dimensionless charge ($\kappa=2$ for a Cooper pair) and
\begin{eqnarray}
Q_{\mu} = \sum_i \int_{\rm Q_i} d\tau {dq_{\mu}^{(i)}(\tau) \over d\tau} \ \delta^3 \left( x- q^{(i)}(\tau) \right) \ ,
\nonumber \\
M_{\mu} = \sum_i \int_{\rm M_i} d\tau {dm_{\mu}^{(i)}(\tau) \over d\tau} \ \delta^3 \left(  x- m^{(i)}(\tau) \right) \ ,
\label{topex2D}
\end{eqnarray}
with $\{ Q_i \}$ and $\{ M_i\}$ representing the world-lines, parametrized by $q^{(i)}$ and $m^{(i)}$, of elementary charges and vortices, respectively (we use natural units $c=1$, $\hbar =1$).
Integrating out the gauge fields (one needs an intermediate regulator for this) gives 
\begin{equation}
S_{\rm linking} = 2\pi i \ \int d^3 x \ Q_{\mu} \epsilon_{\mu \alpha \nu} {\partial_{\alpha} \over -\nabla^2} M_{\nu} \ .
\label{linking}
\end{equation}
For charge-anticharge and vortex-antivortex fluctuations, represented by closed loops $\{ Q_i \}$ and $\{ M_i\}$, this is the sum of the integer Gauss linking numbers between closed loops of the two kinds. These linking numbers represent the Aharonov-Bohm/Aharonov-Casher phases accumulated when one charge completely encircles a vortex and viceversa. Because of the factor $(2\pi i)$ such integer linking numbers do not contribute to the partition function. They do, however for generic, infinitely extended world-lines of charges and vortices. The action (\ref{CS}) is the local representation of these topological interactions.

The charge and vortex number currents $Q_{\mu}$ and $M{\mu}$ are conserved. Correspondingly, the gauge fields are invariant under the U(1) gauge transformations $a_{\mu} \to a_{\mu} + \partial_{\mu} \lambda $ and $b_{\mu} \to b_{\mu} + \partial_{\mu} \chi $. The full effective action for the SIT must then respect these two gauge invariances. The Chern-Simons term is the only marginal gauge invariant term in 2D since it is the unique gauge invariant term involving only one field derivative. Topological interactions thus dominate near the SIT. From a purely field-theoretic point of view, the charge and vortex world-lines represent the singularities in the dual field strengths $f_{\mu} = \epsilon_{\mu \alpha \nu} \partial_{\alpha} b_{\mu}$ and $g_{\mu} = \epsilon_{\mu \alpha \nu} \partial_{\alpha} a_{\mu}$ arising from the compactness of the two U(1) gauge groups \cite{polyakov}. A proper formulation of a compact U(1) gauge theory, however, requires the introduction of an ultraviolet lattice regularization \cite{polyakov}. This was done for the mixed Chern-Simons model in \cite{dst}. 

It is now easy to derive what the possible phases near the SIT are. If only charges condense, the current $Q_{\mu}$ becomes a field that can be expressed as $Q_{\mu} = (\sqrt{\kappa} / 2\pi) \epsilon_{\mu \alpha \nu} \partial_{\alpha} c_{\nu}$, while $M_{\mu}$ vanishes.  The field $c_{\mu}$ can then be reabsorbed by a shift of $b_{\mu}$. If we couple the charge current to a probe electromagnetic gauge field $A_{\mu}$ we obtain the action 
\begin{equation}
S = \int d^3 x \ i  {\kappa \over 2\pi} a_{\mu} \epsilon_{\mu \alpha \nu} \partial_{\alpha} d_{\nu}
+ \  i  {\kappa \over 2\pi} b_{\mu} F_{\mu} \ ,
\label{probesc}
\end{equation}
where $d_{\mu} = b_{\mu} + c_{\mu}$ and $F_{\mu} = \epsilon_{\mu \alpha \nu} \partial_{\alpha} A_{\nu}$ is the dual field strength. This shows that the Chern-Simons term, that must be integrated over $a_{\mu}$ and $d_{\mu}$, decouples. The integration over $b_{\mu}$, instead, yields the electromagnetic effective action. To do it, we need a gauge invariant regulator, which, to dominant order, must be constructed from the ``electric" and ``magnetic" fields of the $b_{\mu}$ gauge potential. This gives 
\begin{equation}
S_{\rm eff}^{\rs SC} \propto \int d^3 x \ v F_0 {1\over -\nabla_3^2} F_0 + {1\over v} F_i  {1\over -\nabla_3^2} F_i \ ,
\label{effsc}
\end{equation}
where $\nabla_3 = \partial_0 \partial_0 + v^2 \nabla_2$ and $v$ is the speed of light in the medium. In Coulomb gauge $A_0=0$, $\partial_i A_i=0$, the effective action reduces to 
\begin{equation}
S_{\rm eff}^{\rs SC} \propto \int d^3 x \ A_iA_i \ ,
\label{effsccoulomb}
\end{equation}
and the induced current $j_i = \delta S_{\rm eff}/\delta A_i$ satisfies the London equations,
\begin{eqnarray}
\partial_0 {\bf j} &\propto& {\bf E} \ ,
\nonumber \\
{\rm rot} \ {\bf j} &\propto& B \ .
\label{london}
\end{eqnarray}
This implies that the electric condensation phase is a superconductor. 

If only vortices condense, the effective action is derived by setting $Q_{\mu} =0$ and coupling, as before, a probe electromagnetic gauge field to the charge current,
\begin{equation} 
S =  \int d^3x \ i  {\kappa \over 2\pi} a_{\mu} \epsilon_{\mu \alpha \nu} \partial_{\alpha} b_{\nu} +i{\sqrt{\kappa}\over 2\pi}
b_{\mu} \left( \kappa e F_{\mu} + 2\pi M_{\mu} \right) \ .
\label{probesi}
\end{equation}
As before, we need gauge invariant regulators to obtain the effective action, but for both the gauge fields $a_{\mu}$ and $b_{\mu}$ this time. Since the Chern-Simons term survives in this case, both fictitious gauge fields are massive. After removing the regulator we obtain the local effective action
\begin{equation}
S_{\rm eff}^{\rs SI} \propto \int d^3 x  \ v \left(\kappa e F_0 + 2\pi M_0 \right)^2 + {1\over v} \left( \kappa e F_i +2\pi M_i \right)^2\ .
\label{effsi}
\end{equation}
which is nothing else but the non-relativistic version of the Polyakov compact QED action\,\cite{polyakov} (a completely rigorous formulation requires, of course, the introduction of a lattice regularization). An important point here is that, in the vortex condensate, the vortex number is not conserved. This is reflected by the presence of instantons $M = \partial_0 M_0 + \partial_i M_i$ at the end of the vortex world lines $M_{\mu}$\,\cite{polyakov}. These instantons force the electric flux into strings, dual images of Abrikosov fluxes, that cause linear confinement of charges\,\cite{polyakov}, leading to an infinite resistance state, the superinsulator \cite{dst, doniach1998, vinokur2008superinsulator, vinokurannals, dtv}. 

Finally, the state where none of the condensates can form, $Q_{\mu} =0$, $M_{\mu}=0$, is characterized by the purely topological mixed Chern-Simons long distance effective action. This intermediate state, that can appear between the superconductor and the superinsulator, was originally called a quantum or Bose metal\,\cite{qm}. Our approach shows that this intermediate Bose metal is a topological insulator\,\cite{moore}, with the quantum resistance arising exclusively from the conductance along the edges\,\cite{dtlv}. Note, however that, while the flux of this topological insulator is $\pi$, the charge is $2e$ instead of $e$ since it is a Cooper pair state. This bosonic topological insulator is thus a level 1 topological insulator, with no ground state degeneracy on the torus. 

Note that {\it no material or disorder parameter} has entered the above derivations, which are entirely predicated on the topological interactions alone. The material parameters do determine the London penetration depth of the superconductor and the string tension of the superinsulator and, thus, as we now show, the conditions for the particular scenario of the transition between the phases, but they are totally irrelevant as far as the nature of the possible phases is concerned. 

As we have have mentioned above, the next order terms in the derivative expansion of the effective action (\ref{CS}) contain two derivatives: gauge invariance requires then that they must be built from the ``electric" ($f_i$ and $g_i$) and ``magnetic" ($f_0$ and $g_0$) fields of the two gauge potentials. The most general possible gauge invariant action up to two field derivatives is then given by 
\begin{eqnarray}
S &=& \int d^3 x\  i  {\kappa \over 2\pi} a_{\mu} \epsilon_{\mu \alpha \nu} \partial_{\alpha} b_{\nu} 
\nonumber \\
&+& {1\over 2e^2_v \mu_P} f_0 f_0+ {\varepsilon_P \over 2e^2_v} f_i f_i 
+{1 \over 2e^2_q\mu_P} g_0 g_0 + {\varepsilon_P \over 2e^2_q} g_i g_i 
\nonumber \\
&+&i\sqrt{\kappa} a_{\mu} Q_{\mu} +i\sqrt{\kappa} b_{\mu} M_{\mu} \  ,
\label{nonrel}
\end{eqnarray}
with the magnetic permeability $\mu_{\rs P}$ and the electric permittivity $\varepsilon_{\rs P}$\,\cite{dtlv}, which determine the speed of light $v = 1/\sqrt{\mu_{\rs P} \varepsilon_{\rs P}}$ in the material. The two coupling constants $e_q^2$ and $e_v^2$ are phenomenological parameters having the dimensionality of [mass] and comprising the remaining material characteristics relevant to this order. It can be shown\,\cite{dtlv} that $e_q^2=e^2/d$, $e_v^2 = \pi^2 /(e^2\lambda_{\perp})$, with $d$ being the thickness of the 2D film and $\lambda_{\perp} = \lambda_{\rs L}^2/d$ the Pearl length and  $\lambda_{\rs L}$ being the London length of the bulk. This identification, however, is not relevant for the  the structures of phases emerging in the critical vicinity of the SIT. One can simply consider $e_q^2$ and 
$e_v^2$ as phenomenological parameters embodying material parameters and effects of disorder. Note that to this order in the derivative expansion, the effective action is perfectly dual with respect to the mutual exchange of charge and vortex degrees of freedom and the corresponding coupling constants. Possible duality breaking is a higher order effect. In field theory, this duality under the transformation $g= e_v/e_q \to 1/g$ goes under the name of S duality (strong-weak coupling duality). 

Upon addition of the usual electromagnetic action both gauge fields acquire a topological Chern-Simons mass. In the relativistic case $\mu_{\rs P} = \varepsilon_{\rs P} = 1$ this is $m=\kappa e_q e_v/2\pi$ \cite{jackiw}. In the non-relativistic case it is modified to $m = \mu_P \kappa e_q e_v/2\pi$ if the dispersion relation $E = \sqrt{m^2 v^4 + v^2 p^2}$ is used \cite{dtlv}. Note that both the coupling constants $e_q^2$ and $e_v^2$ have dimension [mass]. As a consequence, the quadratic terms are power-counting infrared irrelevant: this is the reason why they do not enter in the determination of the nature of the various phases. Chern-Simons theories, however, are plagued by an anomaly: the ground state wave function differs if the model is considered as a purely topological model or as the limit $m\to \infty$ of a topologically massive theory \cite{djt}. The former is not normalizable, only the latter makes physical sense. This has the consequence that the quadratic term are actually non-perturbatively relevant, since they can drive the system toward fixed points different from the bosonic topological insulator. These correspond to continuous phase transitions \cite{peng}, whose location is determined by an energy-entropy balance for the topological excitations $Q_{\mu}$ and $M_{\mu}$ \cite{dst, dtlv}. The transition can also be a direct first-order transition between a superconductor and a superinsulator. The proper derivation of these results requires an ultraviolet regularization of the model on a scale corresponding to the coherence length $\xi$, for example a lattice of spacing $\ell = \xi$. The resulting phase structure \cite{dst,dtlv} is shown in Fig.\,1. 

\begin{figure}
\includegraphics[width=8cm]{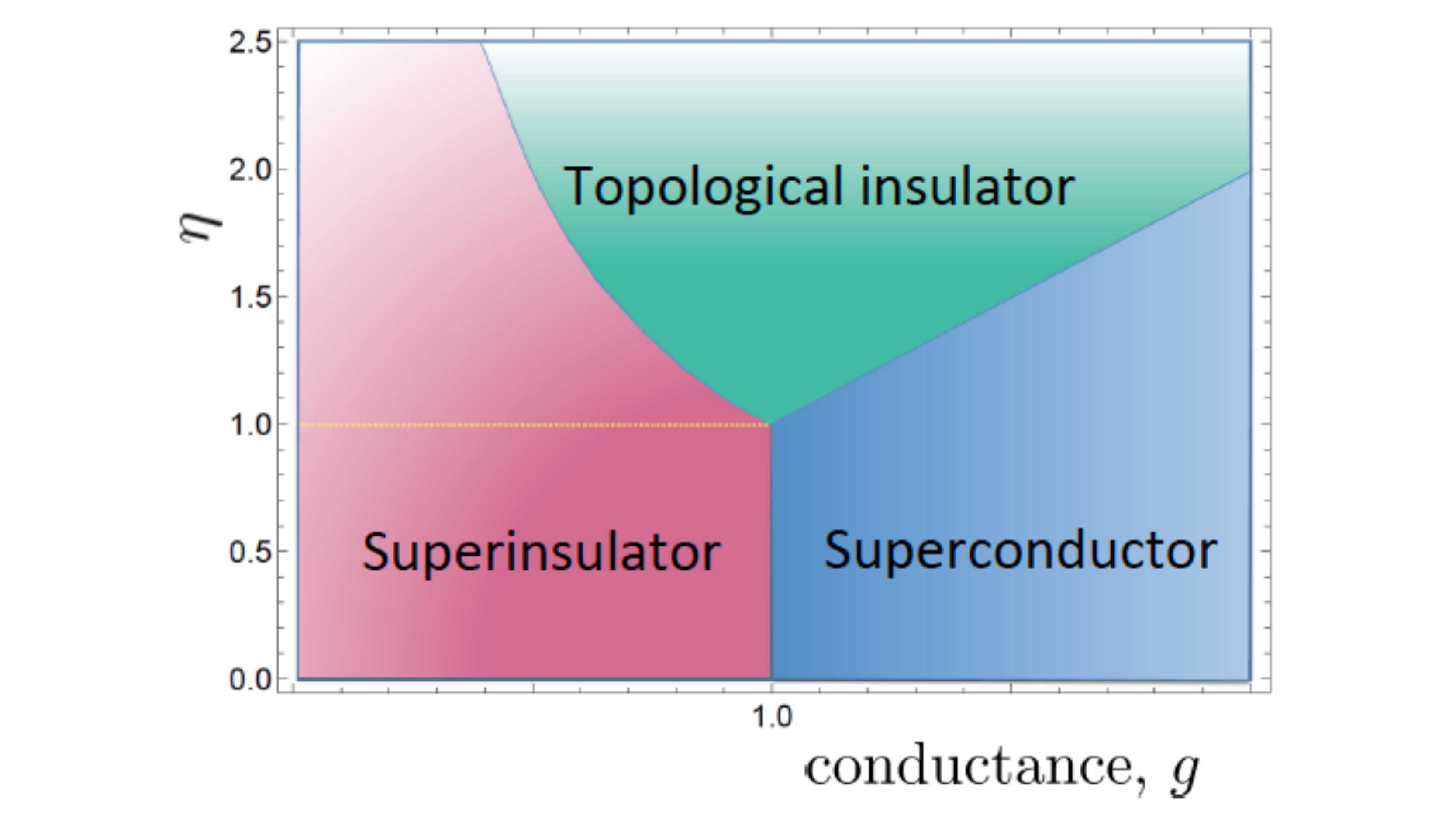}
\caption{\label{fig:Fig.1} {Phase diagram of the vicinity of the SIT.} Tuning the parameter $g=(\pi/e^2) \sqrt{d / \lambda_{\perp}}$, one drives the system across the SIT. The quantity $\eta$ characterizes the strength of quantum fluctuations in a given material. }
\end{figure}

 The two parameters driving the quantum phase structure are the conductance $g=(\pi/e^2) \sqrt{d / \lambda_{\perp}}$ and 
 $\eta = (1/\alpha) f(K, v)$ with $\alpha$ the fine structure constant and $K $ the Landau parameter of the material. The appearance of $\alpha$ in the denominator of $\eta$ shows that the intermediate bosonic topological insulator (Bose metal) phase opens up only if quantum fluctuations are strong enough in a particular material. 
 
The same approach applies also to describe the finite-temperature behaviour of the phases near the SIT\,\cite{dtv}. Specifically, linearly confined charges in the superinsulator are liberated at the deconfinement phase transition,  where the string tension vanishes. In 2D, this deconfinement transition is of the Berezinskii-Kosterlitz-Thoulsess (BKT)\,\cite{ber,KT}\,type \cite{yaffe} and has been recently observed experimentally\,\cite{chargeBKT}. In 3D superinsulators\,\cite{dtv}, instead, the resistance is predicted to have Vogel-Fulcher-Tamman criticality\,\cite{vft}. This critical behaviour has also been recently experimentally observed in InO films\,\cite{shahar}, which have a thickness $d\gg \xi$, making them good candidates for 3D superinsulators. This is confirmed also by the apparent violation of charge-vortex duality in InO films\,\cite{noise}: in 3D, duality is between electric and magnetic fields, not between charges and vortices. A further confirmation of the string confinement picture of superinsulation comes from the recent measurement of strong noise near the threshold voltage\,\cite{noise}. Indeed, the threshold voltage corresponds to the critical strength when an applied voltage starts creating strips of normal insulator, carrying the current, mixed with the superinsulating matrix. The proximity to this dynamic phase transition implies exactly strong current fluctuations near the threshold voltage. 

Finally we conclude by commenting on the suggestions\,\cite{mezard, noise} that many-body localization (MBL)\,\cite{mbl} may be an alternative to the string confinement mechanism for superinsulation. This is not so. Indeed, it has been recently pointed out\,\cite{scardicchio, conf1, conf2} that MBL can arise  {\it without exogenous disorder} due to strong, confining interactions alone, by mechanisms such as the mixing of the charge superselection sectors implied by a gauge symmetry. In the example discussed in\,\cite{scardicchio} this mixing arises in the course of the temporal evolution of quantum states, the mixing mechanism playing effectively the role of a disorder average. This process was identified exactly as a transport-inhibiting mechanism due to confinement in the Schwinger model in 1D. In the present setting, it is the Polyakov monopole instantons that play the role of endogenous, spontaneous disorder. Accordingly, the summation over the instanton gas configurations acts as averaging over disorder as pointed out already in the original literature\,\cite{polyakov}. Importantly, the instanton formulation describes not only 1D, but the 2D and 3D physical dimensions as well. This spontaneous disordering mechanism by instantons has the same effect, that of mixing, in this case, the flux superselection sectors, leading to the survival of only the neutral charge sector as the physical state, while all other, charged states are localized on the string scale. Hence inhibition of the charge transport and the infinite resistance. In the present context, MBL is a different name for the 50-year old phenomenon of confinement and again it is endogenous, external disorder plays no role. 
The same confinement mechanism that prevents the observation of quarks is thus responsible for the absence of charged states and the infinite resistance in superinsulators.

\textit{Acknowledgments--} M. C. D. thanks CERN, where she completed this work, for kind hospitality. The work at Argonne (V.M.V.) was supported by the U.S. Department of Energy, Office of Science, Basic Energy Sciences, Materials Sciences and Engineering Division.

\end{document}